\documentclass[a4paper]{article}
\def\dotFive#1{{\mathop{#1}\limits^{\vbox to-9.5pt{\kern-10pt
 \hbox{\rm...}\vskip-14.5pt\hbox{\rm\null\hskip1.5pt..}\vss}}}}

\def\dotSix#1{{\mathop{#1}\limits^{\vbox to-9.5pt{\kern-10pt
 \hbox{\rm...}\vskip-14.5pt\hbox{\rm\null\hskip0pt...}\vss}}}}

\begin {document}

\title { From a Mechanical Lagrangian to the Schr\"odinger Equation. 
A Modified Version of the Quantum Newton's Law \\ }
\author{A.~Bouda\footnote{Electronic address: 
{\tt bouda\_a@yahoo.fr}} \\
Laboratoire de Physique Th\'eorique, Universit\'e de B\'eja\"\i a,\\ 
Route Targa Ouazemour, 06000 B\'eja\"\i a, Algeria\\}

\date{\today}

\maketitle

\begin{abstract}
\noindent
In the one-dimensional stationary case, we construct a 
mechanical Lagrangian describing the quantum motion of 
a non-relativistic spinless system. This Lagrangian is written 
as a difference between a function $T$, which represents the 
quantum generalization of the kinetic energy and which depends on 
the coordinate $x$ and the temporal derivatives of $x$ up the 
third order, and the classical potential $V(x)$. The Hamiltonian 
is then constructed and the corresponding canonical equations are 
deduced. The function $T$ is first assumed arbitrary. The 
development of $T$ in a power series together with the dimensional 
analysis allow us to fix univocally the series coefficients by 
requiring that the well-known quantum stationary Hamilton-Jacobi 
equation be reproduced. As a consequence of this approach, we 
formulate the law of the quantum motion representing a new version 
of the quantum Newton's law. We also analytically establish the 
famous Bohm's relation 
$\mu \dot{x} = \partial S_0 /\partial x $ 
outside of the framework of the hydrodynamical approach and show 
that the well-known quantum potential, although it is a part of 
the  kinetic term, it plays really a role of an additional 
potential as assumed by Bohm. 
\end{abstract}

\vskip\baselineskip



\newpage

\vskip0.5\baselineskip
\noindent
{\bf 1\ \ Introduction }
\vskip0.5\baselineskip

In the one-dimensional space, the quantum stationary Hamilton-Jacobi 
equation (QSHJE) for a non-relativistic spinless particle of 
mass $\mu$ and energy $E$ is \cite{Messiah} 
\begin {equation}
{1\over 2 \mu} \left({\partial S_0 \over \partial x}\right)^2 + V(x)-E= 
{\hbar^2\over 4\mu}  \left[{3\over 2}\left( 
{\partial S_0 \over\partial x}\right)
^{- 2 }\left({\partial^2 S_0 \over \partial x^2}\right)^2-
\left( {\partial S_0 \over \partial x  }\right)^{- 1 }
\left({\partial^3 S_0 \over \partial x^3  }\right) \right] \; ,
\end {equation}
where $S_0$ and $V(x)$ are respectively the reduced action and 
the external potential. The solution of this equation, investigated 
in \cite{Fl86,FM1,FM2,PCTR}, is written in \cite{QNL} as  
\begin {equation}
S_0=\hbar \arctan\left(a{\phi_1 \over \phi_2}+b\right)
+\hbar \kappa \; ,
\end {equation}
where $(\phi_1,\phi_2)$ is a real set of independent solutions 
of the Schr\"odinger equation (SE) and $(a, b, \kappa)$ are real 
integration constants satisfying the condition $a \ne 0$.

In Bohm's theory \cite{Bohm}, Eq. (1) can be obtained from 
the SE by writing the wave function in the form
\begin {equation}
\phi(x)= A(x) \ \exp\left({i\over\hbar }S_0(x)\right) \; .
\end {equation} 
It is shown that the real function $A(x)$ is proportional to 
$(\partial S_0 /\partial x)^{-1/2}$ \cite{Messiah, PCTR}.

In one dimension, the well-known quantum potential, represented 
by the term proportional to $\hbar^2$ in (1), is written in terms 
of higher derivatives of $S_0$. Thus, it seems not correct to 
consider in a dynamical equation this term as a potential but it 
may be a quantum correction of the kinetic part represented by 
the first term in (1). However, we will show in Section 5 that 
this term plays indeed a role of a an additional potential.

An unsatisfactory feature of the form (3) of the wave function 
is the fact that for bound states, described by real wave functions
up to a constant phase factor, $S_0$ must be constant. Floyd remarked  
this problem and proposed to use a trigonometric representation 
in the real wave function cases \cite{Fl1, Fl2}. He also proposed 
that quantum trajectories were obtained by using Jacobi's 
theorem \cite{Fl2,EV},
\begin {equation}
t-t_0={\partial S_0\over\partial E} \; , 
\end {equation}
as in classical mechanics. In Ref. \cite{reply}, it is shown
that the resulting trajectories depend on the choice of the couple 
of solutions of the SE used in the expression of $S_0$. This 
represents an unsatisfactory feature since the mathematical choices 
affect the physical results. Furthermore, in Ref. \cite{TCQNL}, it 
is shown that the time delay in tunneling through a potential 
barrier, calculated by using Jacobi's theorem as proposed by Floyd, 
manifests also some ambiguities.  

From an equivalence postulate, Faraggi and Matone \cite{FM1,FM2} 
derived quantum mechanics. They deduced that the wave function  
must be written in the unified form 
\begin {equation}
\phi(x)= \left({\partial S_0 \over \partial x }\right)^{-1/2}
\left[\alpha\ \exp\left({i\over\hbar }S_0\right)
+\beta\ \exp\left(-{i\over\hbar }S_0\right)\right]
\end {equation} 
both for bound and unbound states, $\alpha$ and  $\beta$ being 
complex constants. In the case where the wave function $\phi$ 
is real, we have $|\alpha|=|\beta|$ but never $S_0=cte$.
This result is also reproduced in Ref. \cite{PCTR} outside of    
the framework of the equivalence postulate. 

Recently, by appealing to the quantum transformation 
\cite{FM2,FM3} which allows to write the QSHJE in the classical 
form, the relation
\begin {equation}
\dot{x}{\partial S_0 \over \partial x} = 2[(E-V(x)]  
\end {equation}
is derived in Ref. \cite{QNL}. It leads to a third order differential 
equation which appears as a quantum generalization of the first 
integral of Newton's law \cite{QNL}. The relativistic version of 
this law is also obtained for a spinless particle in Ref. 
\cite{RQNL}. When the quantum coordinate \cite{FM2,FM3} is used 
to apply Jacobi's theorem \cite{QNL} or to express the Lagrangian 
in order to obtain the equation of motion \cite{reply}, the 
formalism does not seem to suffer from any mathematical ambiguity. 
Furthermore, the fundamental result is reproduced with many 
formulations \cite{reply}. However, some unsatisfactory features 
are appeared. First, at the turning points (where $V(x)=E$), 
since $\partial S_0 / \partial x$ never has a vanishing value, 
relation (6) indicates that $\dot{x}=0$. Then, from (6) we can 
show that all the higher temporal derivatives of $x$ take a vanishing 
value at the turning points: 
$\dot{x}=0$, $\ddot{x}=0$, $\dot{\ddot{x}}=0$, 
... This means that when the particle gets to these points, it will 
stand still forever. Another feature which seems to be unsatisfactory 
is the extreme rapid divergence of the velocity in the classically 
forbidden regions which is manifested for the three different 
potentials considered in \cite{TCQNL}. Other comments about 
relation (6) are given by Floyd in \cite{comm}.

In this paper, we present a new version of the quantum law of 
motion free from the previous criticisms. In Section 2, we propose 
a generalization of the classical mechanics to the cases where the 
Lagrangian is depending on $x$, $\dot{x}$, $\ddot{x}$ and 
$\dot{\ddot{x}}$. In Section 3, we will use a dimensional analysis 
to construct a quantum Lagrangian for a spinless particle. We will 
show in Section 4 that the well-known QSHJE can be reproduced from 
the previous Lagrangian with a judicious choice of some parameters. 
We, therefore, present the resulting quantum law of motion and 
apply our result in the free particle case.

\vskip0.5\baselineskip
\noindent
{\bf 2\ \ The Generalized Classical Mechanics }
\vskip0.5\baselineskip

Let us consider any Lagrangian, $L$, depending on 
$(x, \dot{x},\ddot{x},\dot{\ddot{x}}, t)$. In the next section 
we will justify the choice of this set of variables. Let us define 
the Hamilton's principal function as
\begin {equation}
S=\int L(x, \dot{x},\ddot{x},\dot{\ddot{x}}, t)\ dt  \; .
\end {equation}
The least action principle leads to the generalized Euler-Lagrange 
equation 
\begin {equation}
{d^3 \over dt^3}{\partial L \over \partial \dot{\ddot{x}}}- 
{d^2 \over dt^2}{\partial L \over \partial \ddot{x}}+{d \over dt}
{\partial L \over \partial \dot{x}}-{\partial L \over \partial x}=0\;.  
\end {equation}
The  corresponding Hamiltonian is constructed in Ref. \cite{reply}. 
Its expression is given by 
\begin {eqnarray}
H=\left({\partial L \over \partial \dot{x}}-{d \over dt}{\partial L 
\over \partial \ddot{x}}+{d^2 \over dt^2}{\partial L \over \partial 
\dot{\ddot{x}}} \right) \dot{x} \hskip15mm&& \nonumber\\
+ \left({\partial L \over \partial \ddot{x}}-{d \over dt}
{\partial L \over \partial \dot{\ddot{x}}}\right) \ddot{x}+
{\partial L \over \partial \dot{\ddot{x}}}\dot{\ddot{x}}-L\; ,
\end {eqnarray}
so that if we require that $\partial L / \partial t=0$, with the use 
of (8), we obtain ${dH / dt}=0$. This means that $H$ is a 
constant of motion when the time is considered uniform.

The conjugate momentum, $P$, must be defined so that if we require 
that the space be homogeneous, $P$ must be a constant of motion.
In other words, if we require that ${\partial L / \partial x}=0$,
we must obtain $\dot{P} \equiv dP/dt = 0$. With the use of Eq. (8), 
it follows that 
\begin {equation}
P={\partial L \over \partial \dot{x}}-{d \over dt}{\partial L 
\over \partial \ddot{x}}+{d^2 \over dt^2}{\partial L \over 
\partial \dot{\ddot{x}}}\; .
\end {equation}
Thus, the generalized Euler-Lagrange equation takes the form
\begin {equation}
\dot{P} = {\partial L \over \partial x} \; .
\end {equation}
We will call $P$ the principal conjugate momentum. The form (9) 
of the Hamiltonian suggests that we define two secondary 
conjugate momentums
\begin {equation}
\Pi = {\partial L \over \partial \ddot{x}} - 
{d \over dt}{\partial L \over 
\partial \dot{\ddot{x}}} 
\end {equation}
and
\begin {equation}
\Xi ={\partial L \over \partial \dot{\ddot{x}}}  \; ,
\end {equation}
so that 
\begin {equation}
H = P \dot{x} + \Pi \ddot{x} + \Xi \dot{\ddot{x}} -L  \; .
\end {equation}
By taking into account relations (10), (11), (12) and (13), from 
(14) we deduce that 
\begin {equation}
dH = - \dot{P} dx + \dot{x} dP - \dot{\Pi} d\dot{x} +
\ddot{x} d\Pi - \dot{\Xi} d\ddot{x} + \dot{\ddot{x}} d\Xi - 
{ \partial L \over \partial t } dt \; .
\end {equation}
If we suppose that the Hamiltonian can be written as follows 
\begin {equation}
H = H(x,P; \dot{x}, \Pi; \ddot{x} , \Xi;t)   \; ,
\end {equation}
we therefore deduce that
\begin {equation}
dH = { \partial H \over \partial x} dx + 
     { \partial H \over \partial P} dP +
     { \partial H \over \partial \dot{x}} d\dot{x} +
     { \partial H \over \partial \Pi} d\Pi + 
     { \partial H \over \partial \ddot{x}} d\ddot{x} + 
     { \partial H \over \partial \Xi} d\Xi + 
     { \partial H \over \partial t } dt \; .
\end {equation}
Identifying (15) and (17), we obtain the generalized 
canonical equations
\begin {eqnarray}
\dot{x} & = &  { \partial H \over \partial P} \; , \\
\dot{P} & = & -{ \partial H \over \partial x}  \; ,\\
\ddot{x} & = & { \partial H \over \partial \Pi} \; , \\
\dot{\Pi} & = & -{ \partial H \over \partial \dot{x}} \; , \\
\dot{\ddot{x}} & = & { \partial H \over \partial \Xi} \; ,\\
\dot{\Xi} & = & -{ \partial H \over \partial \ddot{x}} \; ,\\ 
{ \partial L \over \partial t } & = & -{ \partial H \over \partial t }\; .
\end {eqnarray}
We would like to indicate that the analogue of these equations 
for a Lagrangian depending only on $(x,\dot{x},\ddot{x})$ is obtained 
in Ref. \cite{Salesi}.

We stress to call the reader attention to a contradiction which 
seems to appear when we compare the Lagrangian formulation and the 
Hamiltonian one. Generally, relation (8) is a six order differential 
equation while the last set of equations seems to lead to a fourth 
order equation. In fact, since $H$ is a function of 
$(x,P,\dot{x},\Pi,\ddot{x},\Xi, t)$, by combining (18), (20) and (22), 
we would be able to express $P$, $\Pi$ and $\Xi$ in terms of $x$, 
$\dot{x}$, $\ddot{x}$, $\dot{\ddot{x}}$ and $t$. Then, Eq. (19) 
would be written as a fourth order differential equation, expressed 
only in terms of $x(t)$ and its temporal derivatives. However, as 
we will see in Appendix II where a concrete case is examined, 
Eqs. (18) and (20) are identities and can not be combined with 
(22) in order to express $(P,\Pi,\Xi)$ in terms of 
$(x,\dot{x},\ddot{x},\dot{\ddot{x}}, t)$. So, there is no 
contradiction. We would like to add that a similar false 
contradiction appears also when we consider a Lagrangian 
depending only on $(x,\dot{x},\ddot{x}, t)$.

\vskip0.5\baselineskip
\noindent
{\bf 3\ \ The Quantum Lagrangian}
\vskip0.5\baselineskip

Our goal is to construct in the stationary case a Lagrangian 
from which we can start to get to the well-known QSHJE, which 
in turn, leads to the SE. The higher derivatives of $S_0$ 
appearing in (1) suggest that our Lagrangian depends on $x$, 
$\dot{x}$ and higher temporal derivatives of $x$. Since the 
only parameters which appear in the SE as well as in the QSHJE 
are the mass $\mu$ and the Planck constant $\hbar$, the 
Lagrangian must also depend on $\mu$ and $\hbar$
\[
L = L(\mu, \hbar, x, \dot{x}, \ddot{x}, ... )   \; .
\]
Of course, we have not taken into account the possibility of the 
dependence on other physical parameters through the external 
potential. We will see that this fact is not important for our 
construction.

The reduced action, given in (2), depends on the energy $E$ through 
the functions $\phi_1$ and $\phi_2$ and on two non-additive 
integration constants $a$ and $b$. This indicates that the 
fundamental law describing the quantum motion is a fourth order 
differential equation. It follows that $x(t)$ will contain four 
integration constants as it is in the earlier formulations of 
trajectory representation of quantum mechanics \cite{QNL,EV,Fl4}.
From the mathematical point of view, Eq. (8) indicates that it is 
sufficient to write the Lagrangian as a function of 
$(x,\dot{x},\ddot{x})$ to obtain a fourth order differential 
equation. Nevertheless, from the physical point of view, the set 
$(x,\dot{x},\ddot{x})$ is not sufficient to define the 
``mechanical state'' of the system. In fact, the knowledge of 
this set at any time does not allow to determine the future 
positions of the system since the fundamental law of motion 
must be a fourth order one. However, the  knowledge at any time 
of the set $(x,\dot{x},\ddot{x},\dot{\ddot{x}})$ is sufficient 
to predict the future motion since it must allow to determine 
the four integration constants. Thus, the ``mechanical state'' 
is defined by this last set and the Lagrangian must be 
written as 
\begin {equation}
L = L(\mu, \hbar, x, \dot{x}, \ddot{x}, \dot{\ddot{x}})   \; .
\end {equation}
Note that the analogue of this reasoning for classical 
mechanics was proposed by Landau-Lifchitz \cite{Landau}. 

Now, one mathematical difficulty appears since Eq. (8) 
indicates that the Lagrangian (25) often leads to a six order 
differential equation. In what follows, we will see how to 
overcome this problem. For the moment, let us write the Lagrangian 
in a natural form
\begin {equation}
L = T(\mu, \hbar, x, \dot{x}, \ddot{x}, \dot{\ddot{x}}) - V(x)  \; .
\end {equation}
where $T$ is the quantum generalization of the kinetic energy 
whose form is assumed independent on the external potential $V(x)$. 
Let us suppose that $T$ is a regular function so that we can 
develop it in a power series with respect to $\hbar$ as follows
\begin {equation}
T = \sum_{n} \hbar^{n} 
    T_{n}(\mu, x, \dot{x}, \ddot{x}, \dot{\ddot{x}})   \; .
\end {equation}
In the limit $\hbar \to 0$, $T$ must not diverge. 
Thus, we impose the condition
\begin {equation}
T_{n} = 0, \ \ \ \ \ \ \ \forall \ \  n<0  \; .
\end {equation}

Before going further, it is interesting to remark that 
in order to simplify our investigation, there are two physical 
conditions that we can impose:

\noindent
- in the limit $\hbar \to 0$, the function $T$ goes to the 
classical expression $\mu \dot{x}^2 /2 $;

\noindent 
- in the absence of the external potential, the space must be
homogeneous, and then the condition 
$\partial L / \partial x = 0$ implies that 
$\partial T / \partial x = 0$, meaning that we can suppress the 
dependence on $x$ of $T$ in (26) and (27).

\noindent 
However, in order to persist in the most general construction 
for $T$ and $L$, we do not impose these two conditions. In what 
follows, we will see that the homogeneity of the space when 
$V(x)=0$ and the relation 
$lim_{\hbar \to 0} \ T = \mu \dot{x}^2 /2 $ 
are consequences of the well-known QSHJE.

Because all the terms of the series (27) must have a dimension 
of the energy, the unit of measurement of the function $T_n$ is 
\begin {equation}
[T_{n}] =  kg^{-n+1}m^{-2n+2}s^{n-2}  \; .
\end {equation}
In relation (27), we see that the only physical parameter 
which may appear in the expression of $T_n$  is the mass 
$\mu$. Then, relation (29) indicates that $\forall \ n \geq 0$ 
the function $T_{n}$ takes the form
\begin {equation}
T_{n} =  {1 \over \mu^{n-1}} f_{n}(x, \dot{x}, \ddot{x}, 
         \dot{\ddot{x}})   \; .
\end {equation}
By taking into account relations (28) and (30), (27) turns out 
to be
\begin {equation}
T =  \sum_{n=0}^{\infty}{\hbar^{n} \over 
    \mu^{n-1}} f_{n}(x, \dot{x}, \ddot{x}, 
         \dot{\ddot{x}})   \; ,
\end {equation}
where the unit measurement of the function $f_{n}$ is
\begin {equation}
[f_{n}] = m^{-2n+2}s^{n-2}  \; .
\end {equation}
As indicated above, the dependence on $\dot{\ddot{x}}$ will 
induce a differential equation of sixth order. The unique manner 
to avoid this difficulty is to assume that this dependence is 
linear. Then, the most general form for $f_n$ is 
\begin {equation}
f_{n}(x, \dot{x}, \ddot{x}, \dot{\ddot{x}}) =  
       u_{n}(x, \dot{x}, \ddot{x}) 
     + \dot{\ddot{x}} \; v_{n}(x, \dot{x}, \ddot{x})   \; ,
\end {equation}
where $u_{n}$ and $v_{n}$ are two functions depending only 
on $x$, $\dot{x}$ and $\ddot{x}$. We mention that the term 
$\dot{\ddot{x}} v_{n}(x, \dot{x}, \ddot{x})$ will induce two 
fifth order terms when we apply (8). However, these terms 
cancel each other out. 

By taking into account relation (32), the unit measurements of 
$u_{n}$ and $v_{n}$ are
\begin {equation}
[u_{n}] = m^{-2n+2}s^{n-2}  \; ,
\end {equation}
\begin {equation}
[v_{n}] = m^{-2n+1}s^{n+1}  \; .
\end {equation}
Assuming $u_{n}$ and $v_{n}$ as regular functions, we 
can develop them in a power series as follows  
\begin {equation}
u_{n} = \sum_{ijk} \alpha_{ijk}^{(n)} x^{k} \dot{x}^{i} \ddot{x}^{j}  \; ,
\end {equation}
\begin {equation}
v_{n} = \sum_{ijk} \beta_{ijk}^{(n)} x^{k} \dot{x}^{i} \ddot{x}^{j} \; ,
\end {equation}
where $\alpha_{ijk}^{(n)}$ and $\beta_{ijk}^{(n)}$ are 
dimensionless real numbers since $u_n$ and $v_n$ 
do not depend on any physical parameter. 

Relations (34) and (36) imply that
$$
i+j+k = -2n+2, \ \ \ \ \ \ \ \ \ \ \ \ -i-2j = n-2 \; , \nonumber
$$
leading to $i=-3n-2k+2$ and $j=n+k$. It follows that all the numbers 
$\alpha_{ijk}^{(n)}$ for which $i \ne -3n-2k+2$ and $j \ne n+k$ must 
take a vanishing value. So, by defining
\begin {equation}
\alpha_{nk} \equiv \alpha_{-3n-2k+2,n+k,k}^{(n)}  \; ,
\end {equation}
we have
\begin {equation}
u_{n} = \sum_{k} \alpha_{nk} 
        {x^k  \ddot{x}^{n+k} \over \dot{x}^{3n+2k-2}} \; .
\end {equation}

Relations (35) and (37) imply that
$$
i+j+k = -2n+1, \ \ \ \ \ \ \ \ \ \ \ \ -i-2j = n+1 \; , \nonumber
$$
leading to $i=-3n-2k+3$ and $j=n+k-2$. It follows that all the 
numbers $\beta_{ijk}^{(n)}$ for which $i \ne -3n-2k+3$ and 
$j \ne n+k-2$ must take a vanishing value. So, by defining
\begin {equation}
\beta_{nk} \equiv \beta_{-3n-2k+3,n+k-2,k}^{(n)}  \; ,
\end {equation}
we have
\begin {equation}
v_{n} = \sum_{k} \beta_{nk} 
        { x^k \ddot{x}^{n+k-2} \over \dot{x}^{3n+2k-3} }\; .
\end {equation}

From (39) and (41), we see that for $k<0$ the functions $u_n$ and 
$v_n$ diverge at $x=0$. In order to avoid these divergences, 
$\alpha_{nk}$ and $\beta_{nk}$ must take a vanishing value for $k<0$. 
Concerning the possible divergences which will appear in the case 
where $\dot{x}$ or $\ddot{x}$ take a vanishing value, in the 
following Sections, we will see that $\dot{x}$ never reaches a 
vanishing value and  $\ddot{x}$ is never present in the denominator. 
Thus, by taking into account relations (33), (39) and (41), 
expression (31) turns out to be
\begin {equation}
T = \sum_{n=0}^{\infty} \sum_{k=0}^{\infty} { \; \hbar^{n} \over \mu^{n-1}} 
    \left[
         \alpha_{nk} {x^k \ddot{x}^{n+k} \over \dot{x}^{3n+2k-2}} 
         + \beta_{nk} {x^k \ddot{x}^{n+k-2} \dot{\ddot{x}} 
           \over \dot{x}^{3n+2k-3}}  
    \right]    \; .
\end {equation}
In this relation, $k$ is considered as an integer number 
meaning that $u_n$ and $v_n$ are assumed infinitely differentiable 
at $x=0$. If it is not the case, in order to keep $k$ integer, 
it is sufficient to substitute in the above relations $x$ by 
$(x-x_0)$, $x_0$ being a point chosen in such a way as to 
have $u_n$ and $v_n$ infinitely differentiable at this point. 

However, with the form (42) of $T$, we have a  problem 
concerning the Hamiltonian formulation. In fact, 
by applying (13), we get an expression for $\Xi$ only in 
terms of $x$, $\dot{x}$ and $\ddot{x}$. This means that $\Xi$, $x$,   
$\dot{x}$ and $\ddot{x}$ can not be considered as independent 
variables in the Hamiltonian approach. In addition, if we apply 
(12) and (20), we see that there is a contradiction 
between the obtained results unless we put $\beta_{nk} =0$ 
for every $n$ and $k$. This forces us to lose the dependence on 
$\dot{\ddot{x}}$. However, as explained at the 
beginning of this Section, in order to obtain a Lagrangian 
describing the ``mechanical state'' of the system, the presence of 
$\dot{\ddot{x}}$ is required. We stress to indicate that it is 
not the generalized classical mechanics presented in Section 2 
which is ambiguous, but the linear terms pose also problems in the 
formulation where we have a Lagrangian of classical type. 
As shown in Appendix I, it is instructive to consider a Lagrangian 
depending only on $x$ and $\dot{x}$ as in classical mechanics, to 
see why linear terms induce mathematical ambiguities and how to 
overcome these difficulties. The solution consists in 
adding to the Lagrangian a quadratic term in $\dot{x}$, 
proportional to a constant $\lambda$, and in taking at the end the 
limit $\lambda \to 0$. That's what we will do in the quantum case. 
We will add a quadratic term in $\dot{\ddot{x}}$ and, with the use 
of (26) and (42), we write the quantum Lagrangian as
\begin {equation}
L = \sum_{n=0}^{\infty} \sum_{k=0}^{\infty} { \; \hbar^{n} \over \mu^{n-1}} 
    \left[
         \alpha_{nk} {x^k \ddot{x}^{n+k} \over \dot{x}^{3n+2k-2}} 
         + \beta_{nk} {x^k \ddot{x}^{n+k-2} \dot{\ddot{x}} 
           \over \dot{x}^{3n+2k-3}}  
    \right]     
     +{1 \over 2} \lambda  \dot{\ddot{x}}^2 - V(x) \; ,
\end {equation}
$\lambda$ being a constant extremely small so that, after having 
obtained the equation of motion, we can take the limit 
$\lambda \to 0$. We stress that $\lambda$ is independent on 
$\mu$ and $\hbar$. Of course, the Lagrangian (43) leads to a six 
order differential equation. However, as we will see in Section 4, 
when we take the limit $\lambda \to 0$, we obtain a fourth order 
equation of motion. We emphasize that this additional term is 
useless in the Lagrangian approach but it is necessary for  
a coherent formulation of the Hamiltonian approach.

Now, let us calculate the conjugate momentums by applying (10), 
(12) and (13). We get
\begin {eqnarray}
P  & = &  \sum_{n=0}^{\infty} \sum_{k=0}^{\infty} 
          {\hbar^{n} \over \mu^{n-1}} 
          \left\{ \phantom{{\dot{x}^n \over \dot{x}^3} } 
          \right. \nonumber \\
   &  &  \left[ \hskip-4.5mm \phantom{Q \over M}
       (3n+2k-2)(n+k-1)\alpha_{nk} +(3n+2k-2)(3n+2k-3) \beta_{nk}
         \right. \nonumber \\ 
   &  &    \left.  
      \hskip10mm -(k+1)(n+k+1) \alpha_{n,k+1} - (k+1)(6n+4k-3) \beta_{n,k+1} 
           \right. \nonumber \\ 
   &  &    \left.        
      \hskip45mm +(k+1)(k+2)\beta_{n,k+2}
       \hskip-4.5mm \phantom{Q \over M}  \right]
      {x^k \ddot{x}^{n+k} \over \dot{x}^{3n+2k-1}} \nonumber \\
   &  & \hskip-3.5mm +\left[ \hskip-4.5mm \phantom{Q \over M}
      - (n+k)(n+k-1) \alpha_{nk} - (n+k)(3n+2k-3) \beta_{nk}  
         \right. \nonumber \\ 
   &  &   \left.       
      \hskip35mm + (k+1)(n+k-1) \beta_{n,k+1}
      \hskip-4.5mm \phantom{Q \over M}  \right] 
     \left. {x^k \ddot{x}^{n+k-2} \dot{\ddot{x}} \over \dot{x}^{3n+2k-2}} 
     \right\rbrace \nonumber \\
   &  & \hskip-3.5mm + \hskip0.8mm \lambda \; \dotFive{x} \; , \\
\Pi  & = &  \sum_{n=0}^{\infty} \sum_{k=0}^{\infty} 
            { \hbar^{n} \over \mu^{n-1}} 
     \left[ \hskip-4.5mm \phantom{Q \over M}  
           (n+k) \alpha_{nk} \right. \nonumber \\ 
    &   &   \left.       
      \hskip12mm + (3n+2k-3) \beta_{nk} - (k+1) \beta_{n,k+1} 
      \hskip-4.5mm \phantom{Q \over M}\right]
           {x^k \ddot{x}^{n+k-1} \over \dot{x}^{3n+2k-2}}  
     - \lambda \;\ddot{\ddot{x}} \; , \\
\Xi & =  & \sum_{n=0}^{\infty} \sum_{k=0}^{\infty} 
          { \hbar^{n} \beta_{nk} \over \mu^{n-1}} 
         {x^k \ddot{x}^{n+k-2} \over \dot{x}^{3n+2k-3}}  
      +  \lambda \;\dot{\ddot{x}} \; .
\end {eqnarray}
Since we see in these three relations nine variables 
$(P,\Pi,\Xi, x, \dot{x},\ddot{x},...,\dotFive{x})$, 
it is clear that in the set $(x,P,\dot{x},\Pi,\ddot{x},\Xi)$, 
all the six variables can be considered independent. 
Then, by using (14), (43) and (46), the Hamiltonian 
takes the form
\begin {eqnarray}
H(x,P,\dot{x},\Pi,\ddot{x},\Xi) = P \; \dot{x} + \Pi \; \ddot{x} 
    - \sum_{n=0}^{\infty} \sum_{k=0}^{\infty} 
    { \hbar^{n} \alpha_{nk} \over \mu^{n-1}} 
    {x^k \ddot{x}^{n+k} \over \dot{x}^{3n+2k-2}} 
\hskip20mm&& \nonumber\\  
    + {1 \over 2 \lambda }
    \left[
    \Xi - \sum_{n=0}^{\infty} \sum_{k=0}^{\infty} 
    { \hbar^{n} \beta_{nk} \over \mu^{n-1}} 
    {x^k \ddot{x}^{n+k-2} \over \dot{x}^{3n+2k-3}}
    \right]^2    
    + V(x)\; .
\end {eqnarray}
In Appendix II, the resulting canonical equations are 
deduced and it is shown that they allow to reproduce 
the same expressions (44), (45) and (46) for the 
conjugate momentums $P$, $\Pi$ and $\Xi$ obtained from the 
Lagrangian (43). In particular, it is also shown that the 
Hamiltonian formulation is equivalent to the Lagrangian one 
and both of them lead to the same law of motion for any 
$\lambda$. This law can be easily obtained by using (44) in 
the expression of $\dot{P}$ given in Appendix II and which 
we obtain by applying (19). Calculating the temporal derivative 
of (44), we see that the expression of $\dot{P}$ has no term 
containing $\dotFive{x}$. The term containing $\dotSix{x}$ is 
proportional to $\lambda$. In conclusion, we obtain in the 
Lagrangian formulation as well as in the Hamiltonian one the 
same fourth order equation of motion when $\lambda \to 0$.

\vskip0.5\baselineskip
\noindent
{\bf 4\ \ Toward the Quantum Stationary Hamilton-Jacobi Equation}
\vskip0.5\baselineskip

Now, our task consists in finding the numerical values of the 
dimensionless parameters $\alpha_{nk}$ and $\beta_{nk}$ with which 
we can reproduce the well-known QSHJE. In the stationary case, the 
Hamilton's principal function is related to the reduced action 
by $S = S_0 - Et$. Then, we write
\begin {equation}
dS = dS_0 - Edt\; .
\end {equation}
On the other hand, by using (7) and (14), we can deduce that 
\begin {equation}
dS = Ldt = Pdx + \Pi d\dot{x} + \Xi d\ddot{x} - Edt\; ,
\end {equation}
where the Hamiltonian $H$ is substituted by $E$. Comparing (48) 
and (49), we obtain
\begin {equation}
dS_0 = Pdx + \Pi d\dot{x} + \Xi d\ddot{x} ,
\end {equation}
from which we deduce
\begin {eqnarray}
{dS_0 \over dx} 
      &  =  & P + \Pi {d\dot{x} \over dx} + 
              \Xi {d\ddot{x} \over dx} \nonumber \\ 
      &   = & P + \Pi {\ddot{x} \over \dot{x}}
           + \Xi {\dot{\ddot{x}} \over \dot{x}} \; .
\end {eqnarray}
We indicate that the left hand side in this last equation 
represents the partial derivative $\partial S_0 / \partial x$ 
appearing in (1) and in the abstract. In fact, in (1),  
$S_0$ depends on $x$ and not on any other variable. 
Furthermore, from the solution (2) of (1), we see that $S_0$  
is a function only of $x$ and some integrations constants. 

Thus, the usual conjugate momentum \cite{Fl86,FM2,PCTR,QNL,EV}, 
represented here by the left hand side of (51), differs from the  
principal conjugate momentum $P$ that we have defined in (10). This 
difference can be explicitly calculated from (51). 

Using expression (44), (45) and (46) in the limit $\lambda \to 0$, 
relation (51) turns out to be
\begin {equation}
{dS_0 \over dx} =   
       \sum_{n=0}^{\infty}  \sum_{k=0}^{\infty} 
       { \hbar^{n} \over \mu^{n-1}}
\left[ 
      A_{nk} {x^k \ddot{x}^{n+k} \over \dot{x}^{3n+2k-1}} 
      + B_{nk} {x^k \ddot{x}^{n+k-2} \dot{\ddot{x}} 
                          \over \dot{x}^{3n+2k-2}}
\right]
    \; ,
\end {equation}
where
\begin {eqnarray}
A_{nk} = (3n^2+2k^2+5nk-4n-3k+2) \alpha_{nk} \hskip38mm&& \nonumber \\
         + (3n+2k-1)(3n+2k-3) \beta_{nk} 
         - (k+1)(n+k+1) \alpha_{n,k+1} \hskip5mm&& \nonumber \\
         - 2(k+1)(3n+2k-1) \beta_{n,k+1}
         +(k+1)(k+2) \beta_{n,k+2}
 \; ,
\end {eqnarray}
and
\begin {eqnarray}      
B_{nk} = -(n+k)(n+k-1) \alpha_{nk} - (3n^2+2k^2+5nk
          \hskip28mm&& \nonumber \\
         -3n-3k-1) \beta_{nk} 
         + (k+1)(n+k-1) \beta_{n,k+1}
    \; .
\end {eqnarray}
With the use of (51), Eq. (14) turns out to be
\begin {equation}
H = \dot{x} {dS_0 \over dx} - L \; .
\end {equation}
Substituting in this last relation $H$ by $E$ and using (43)
in the limit $\lambda \to 0$, we deduce that
\begin {equation}
E-V(x) = \dot{x} {dS_0 \over dx}
  - \sum_{n=0}^{\infty} \sum_{k=0}^{\infty} { \; \hbar^{n} \over \mu^{n-1}} 
    \left[
         \alpha_{nk} {x^k \ddot{x}^{n+k} \over \dot{x}^{3n+2k-2}} 
         + \beta_{nk} {x^k \ddot{x}^{n+k-2} \dot{\ddot{x}} 
           \over \dot{x}^{3n+2k-3}}  
    \right]   
    \; .
\end {equation}
By using this last expression, we can rewrite Eq. (1) in the form
\begin {eqnarray}
      \dot{x} \left({dS_0 \over dx}\right)^3 - 
      { 1 \over 2\mu} \left({dS_0 \over dx}\right)^4 
      + { \hbar^{2} \over 4\mu} 
      \left[{3 \over 2} \left({d^2 S_0 \over dx^2}\right)^2 
      -{dS_0 \over dx}{d^3 S_0 \over dx^3 }
      \right] \hskip20mm&& \nonumber\\
  =\left({dS_0 \over dx}\right)^2 
   \left[ 
      \sum_{n=0}^{\infty} \sum_{k=0}^{\infty} 
           { \hbar^{n} \over \mu^{n-1}} 
    \left(
         \alpha_{nk} {x^k \ddot{x}^{n+k} \over \dot{x}^{3n+2k-2}} 
         + \beta_{nk} {x^k \ddot{x}^{n+k-2} \dot{\ddot{x}} 
           \over \dot{x}^{3n+2k-3}}  
    \right)
    \right]   
  \; .
\end {eqnarray}
With the use of (52) in Appendix III, it is shown  that the 
unique physical solution for $\alpha_{nk}$ and $\beta_{nk}$ 
satisfying this last relation is  
\begin {equation} 
    \alpha_{00}= {1 \over 2} \; , \ \ \ \ \ \ 
    \alpha_{20} =  {5 \over 8} \; , \ \ \ \ \ \ 
    \beta_{20} = -{1 \over 4} \; ,
\end {equation}
\begin {equation}
    \beta_{0k} = \alpha_{1k} = \beta_{1k} = 0 \ \ \forall \ \ k \geq 0 \; ,
     \ \ \ \alpha_{0k} = \alpha_{2k} = \beta_{2k} = 0  
     \ \ \ \forall \ \ k \geq 1 \; ,  
\end {equation}
and
\begin {equation} 
     \alpha_{nk} = \beta_{nk} = 0  
     \ \ \forall \ \ n \geq 3 , \ \ \forall \ \ k \geq 0  \; .
\end{equation} 
Thus, with the use of these values in expression 
(52), the Hamiltonian (47) leads straightforwardly, 
when we take  the limit $\lambda \to 0$, 
to the well-known QSHJE given by (1).

It is also interesting to remark that by using (58), (59) 
and (60) in expression (42), we deduce that 
$lim_{\hbar \to 0} \ T = \mu \dot{x}^2 /2$. Thus, the classical 
expression for the kinetic energy is a consequence of the 
substitution of the general form of the Hamiltonian that we have 
constructed in the well-known QSHJE when we take the limit 
$\hbar \to 0$. 

Since there is a unique physical solution for the set 
$\lbrace \alpha_{nk}, \beta_{nk} \rbrace $ and the fact that 
$\beta_{20}$ is different from 0, we conclude that without 
the linear term in $\dot{\ddot{x}}$ in Eq. (33), our 
initial Lagrangian never allows to reach the QSHJE. This  
constitutes the mathematical reason for which we have imposed 
the presence of this term.

\vskip0.5\baselineskip
\noindent
{\bf 5\ \ The Quantum Law of Motion }
\vskip0.5\baselineskip

With the use of (58), (59 and (60), although we obtain 
from (52)
\begin {equation}
{dS_0 \over dx} =  \mu \dot{x} \; ,
\end {equation}
recalling the famous Bohm's relation postulated a half 
century ago \cite{Bohm}, the principal conjugate momentum
can be derived from (44) and takes the form
\begin {equation}
P   =   \mu \dot{x} - {\hbar^{2} \over 4\mu} 
    \left[ 
         2{\ddot{x}^2 \over \dot{x}^{5}} - 
         {\dot{\ddot{x}} \over \dot{x}^{4}} 
    \right] \; .
\end {equation}
From (43), the expression of the Lagrangian $(\lambda \to 0)$ is 
\begin {equation}
L = {1 \over 2}\mu \dot{x}^2 + {\hbar^{2} \over 4 \mu} 
    \left[
        {5 \over 2} {\ddot{x}^2 \over \dot{x}^4}  
        - {\dot{\ddot{x}} \over \dot{x}^{3}}  
    \right] 
    - V(x)\; .
\end {equation}
It is clear that, thanks to the obtained values for 
$\{ \alpha_{nk},\beta_{nk} \}$, the kinetic term does not depend 
on $x$. This means that in the absence of the external potential, 
$V(x)=0$, we have $\partial L / \partial x =0$ and then we deduce 
that the space is homogeneous. This property was not imposed in the 
present formalism but it is a consequence of the QSHJE. We also 
deduce that the principal conjugate momentum, $P$, is a constant 
of motion in the case where $V(x)=0$.  

By applying (9) or (14), the Hamiltonian becomes
\begin {equation}
H = {1 \over 2} \mu \dot{x}^2 - {\hbar^{2} \over 4\mu} 
    \left[ 
         {5 \over 2} {\ddot{x}^2 \over \dot{x}^{4}}  
        -{\dot{\ddot{x}} \over \dot{x}^{3}}
    \right]  
   + V(x)\; .
\end {equation}
First, this expression can also be obtained from the 
QSHJE, Eq. (1), by using (61) and substituting $E$ by $H$. 
Second, it is interesting to remark that if we define 
\begin {equation}
Q =  - {\hbar^{2} \over 4\mu} 
    \left[ 
         {5 \over 2} {\ddot{x}^2 \over \dot{x}^{4}}  
        -{\dot{\ddot{x}} \over \dot{x}^{3}}
    \right]  
   \; ,
\end {equation}
representing the well-known quantum potential in the QSHJE, Eqs. 
(63) and (64) can be written as 
\begin {equation}
L = {1 \over 2} \mu \dot{x}^2 - [Q+V(x)] \;, \ \ \ \ \ \ \ \ 
H = {1 \over 2} \mu \dot{x}^2 + [Q+V(x)] \; .
\end {equation}
These two last relations constitute a proof that $Q$ really 
plays a role of an additional potential, as predicted by 
Bohm \cite{Bohm}, despite it is a part of the kinetic term $T$. 
This is possible because of the particular values we 
have obtained for $\alpha_{20}$ and $\beta_{20}$.

By construction, we have automatically $dH/dt=0$ since we are in 
the stationary case. Thus, by substituting in (64) $H$ by $E$, 
we obtain a first integral of the analogue of the quantum 
Newton's law. Calculating the total derivative with respect to 
$x$ or $t$ of the two members of (64), we get 
\begin {equation}
\mu \ddot{x} + {\hbar^{2} \over  \mu} 
    \left[
        {5 \over 2} {\ddot{x}^3 \over \dot{x}^6}  
        -2 {\ddot{x}\dot{\ddot{x}} \over \dot{x}^{5}} 
        + {1 \over 4} {\ddot{\ddot{x}} \over \dot{x}^4}
    \right] 
    + {dV \over dx} = 0\; ,
\end {equation}
representing the analogue of the quantum Newton's law. It 
can also be obtained by applying (8) or (11). As expected, 
it is a fourth order differential equation. The general 
solution $x(t)$ will contain four integration constants which
can be determined by the knowledge of the initial conditions 
$x(t_0)=x_0$, $\dot{x}(t_0)=\dot{x}_0$, 
$\ddot{x}(t_0)=\ddot{x}_0$ and 
$\dot{\ddot{x}}(t_0)=\dot{\ddot{x}}_0$. 
In the case where the energy $E$ is known, the three first 
conditions are sufficient. In contrast to the law 
established in \cite{QNL}, through (64) we see that there is  
no derivative of the external potential in the first integral 
of (67).

We remark that in all the above relations, if we put 
$\hbar =0 $, we reproduce the classical formulas.

We observe also that relation (61) can be considered as a law 
of motion. In fact, by using (2), we have
\begin {equation}
\mu {dx \over dt}  = {\hbar a W 
                     \over {a^2 \phi_1^2 + (1+b^2)\phi_2^2
                     +2ab \phi_1\phi_2} } \; ,
\end {equation}
where $W=\phi_2 \; d\phi_1/dx-\phi_1 \; d\phi_2/dx$ is a constant 
representing the Wronskian of the two independent solutions 
$(\phi_1,\phi_2)$ of the SE. Relation (68) is a first order 
differential equation in which we see the presence of three 
integration constants: $a$, $b$ and $E$ through $\phi_1$ and 
$\phi_2$. We stress to indicate that Eq. (68) is independent 
on the choice of the couple $(\phi_1,\phi_2)$. In fact, if we 
use another couple $(\theta_1,\theta_2)$ in (2), with the same 
procedure used in Ref. \cite{TCQNL}, we can find two other 
parameters $(\tilde{a},\tilde{b})$, which we must use instead 
of $(a,b)$ in expression (2), in such a way as to guarantee that 
$dS_0 / dx$ remains invariant. 

In Section 3, we have indicated that a problem about some  
divergences may occur in the kinetic term (42). Concerning 
$\ddot{x}$, since $\beta_{00}=\beta_{01}=\beta_{10}=0$, 
$\ddot{x}$ is never present in the denominator. 
With regard to $\dot{x}$, since $a$ and $W$ are both different 
from 0, Eq. (68) indicates that $\dot{x}$ never reaches a 
vanishing value.

As an example, let us consider the free particle case for which 
$V(x)=0$. Choosing for the SE the two following solutions
\begin {equation}
\phi_1 = \sin (kx) \; , \ \ \ \ \ \ \ \ \phi_2 = \cos (kx) \; , 
\end {equation}
where $k=\sqrt{2 \mu E} / \hbar $, Eq. (68) leads to 
the quantum time equation  
\begin {equation}
a\sqrt{2 E \over \mu }(t-t_0) = 
     {a^2 +b^2 +1 \over 2}x + {1 +b^2 - a^2 \over 4k} \sin (2kx) 
     -{ab \over 2k}\cos (2kx) \; . 
\end {equation}
First, if we put $a=1$ and $b=0$, we reproduce the classical 
relation
\begin {equation}
x = \sqrt{2 E \over \mu }(t-t_0)  \; , 
\end {equation}
as it is the case in the earlier formulations \cite{QNL,Fl4} of 
trajectory representation. Second, in contrast to the 
trajectories obtained in \cite{TCQNL}, we have no nodal structure. 
Third, in the classical limit $\hbar \to 0$, as in \cite{Fl4}, 
a residual indeterminacy subsists. However, in this limit, (70) 
differs from (71) only by the proportionality factor between $x$ 
and $(t-t_0)$. For a particular condition on the parameters 
$a$ and $b$, (70) reduces to (71).

In our point of view, in order to obtain a realistic model, 
it is necessary to generalize the present formulation to the 
three-dimensional space.

\vskip0.5\baselineskip
\noindent
{\bf 6\ \ Conclusion }
\vskip0.5\baselineskip

Before concluding, let us summarize the principal steps of the 
present approach.

After having proposed a generalization of the classical  
mechanics by starting from any Lagrangian depending on  
$(x, \dot{x},\ddot{x},\dot{\ddot{x}}, t)$, our goal was to 
reach the SE by reproducing the well-known QSHJE. 

Our task consisted in establishing a fourth order differential 
equation to describe the quantum motion. For this purpose, by 
appealing to the dimensional analysis, we constructed in the 
stationary case a Lagrangian depending on 
$(x, \dot{x},\ddot{x},\dot{\ddot{x}})$ from which we deduced 
a conjugate momentum which was a constant of motion in the 
absence of the external potential. Although we observed 
that a Lagrangian depending only on $(x, \dot{x},\ddot{x})$ was 
sufficient to establish a fourth order law, we indicated that 
in order to obtain a  Lagrangian which described the ``mechanical 
state'' of the particle, it had also to depend on $\dot{\ddot{x}}$. 
Furthermore, without this dependence, from a mathematical point 
of view it was impossible to reach the QSHJE from our initial 
Lagrangian. Nevertheless, this dependence was linear in order to 
guarantee that the resulting law of motion should be a fourth order 
equation. We kept $\ddot{x}$ in the terms proportional to  
$\dot{\ddot{x}}$ because the induced fifth order terms in the 
equation of motion cancel each other out.

However, we remarked that with such a Lagrangian, it was 
not possible to obtain a coherent Hamiltonian formulation. 
This difficulty was surmounted by adding to the Lagrangian a 
quadratic term in $\dot{\ddot{x}}$ proportional to 
one parameter noted $\lambda$ and which was independent on 
$\hbar$ and the mass $\mu$ of the particle. In order to avoid a 
sixth order law, we took the limit $\lambda \to 0$ after the 
equation of motion was obtained. We stress that this 
quadratic term is useless for the Lagrangian formulation. 
In other words, we have the same result if we do not 
add this term or if we add it and then take the limit 
$\lambda \to 0$. However, in the Hamiltonian approach, 
this term is necessary in order to have a coherent 
formulation. We showed that the two formulations, the 
Lagrangian and Hamiltonian ones, are equivalent.

We would like to indicate that it was possible to obtain all 
the results presented here without appealing to the Hamiltonian 
formulation, in particular to the canonical equations. 
It was sufficient to only use expression (9) of the 
Hamiltonian and to consider expressions (10), (12) and (13) 
of $P$, $\Pi$ and $\Xi$ as mathematical beings but not as 
conjugate momentums. With this procedure, we insist that 
the additional quadratic term in $\dot{\ddot{x}}$ proportional 
to $\lambda$ is not required. However, we developed an 
Hamiltonian formulation in order to prove that such an 
approach was possible in the context of the generalized 
classical mechanics, to confirm the result obtained with 
the Lagrangian formulation and especially to justify that 
$P$, $\Pi$ and $\Xi$ are really conjugate momentums. 

The general Lagrangian and Hamiltonian that we constructed 
depend on some dimensionless parameters. We substituted in the 
well-known QSHJE the expression of the Hamiltonian and 
we determined the values of these parameters in such a way as 
to guarantee the validity of the obtained relation.
We showed that there was a unique physical solution.  
The obtained values allowed to reproduce the famous Bohm's 
relation $dS_0/dx = \mu \dot{x}$ and to show that the quantum 
potential, although it was a part of the kinetic term, it 
really played a role of a potential. We also established 
the fundamental law of the quantum motion and applied our result 
in the free particle case.

In addition, we would like to mention that we have not imposed 
in the formalism to the space to be homogeneous in the absence 
of the external potential since we have not excluded the 
possibility of the dependence on $x$ of $T$. We showed that 
this property was a consequence of the substitution in the QSHJE 
of the general expression of the Hamiltonian that we 
constructed and, in contrast to the earlier formulations 
of trajectory representation,  we deduced that the principal 
conjugate momentum was a constant of motion in the absence of the 
external potential. We have not also imposed to the kinetic term 
to have as a limit when $\hbar \to 0$ the classical expression 
$\mu \dot{x}^2/2$. We showed that this condition was a consequence 
of the QSHJE. In our point of view, this constitutes a 
proof that the function $S_0$ appearing in the QSHJE, Eq. (1), 
and which is related to the Schr\"odinger wave function by (5),  
is really a quantum generalization of the classical reduced 
action. 

To conclude, we would like to emphasize that in the 
context of the following hypothesis:

\noindent
- the Hamilton's principal function is represented as an integral 
of a Lagrangian;

\noindent
- the Lagrangian is a difference between a kinetic term, $T$, and 
the external potential $V(x)$;

\noindent
- the kinetic term is a function of 
$(\hbar, \mu, x, \dot{x}, \ddot{x},\dot{\ddot{x}} )$ and its form 
does not depend on $V(x)$;

\noindent
- the resulting equation of motion is a fourth order one,

\noindent
we showed that the Lagrangian which leads to the well-known QSHJE, 
and then to the SE, is unique and is the one given by (63). 
Of course, in the context of the above hypothesis, the resulting 
quantum law of motion, Eq. (67), is also unique. 
%
%
\vskip\baselineskip
\noindent
{\bf Appendix I}
\vskip\baselineskip
%
%

In this Appendix, we would like to explain in the classical 
Lagrangian case why linear terms induce ambiguities and how to 
overcome these difficulties. For this purpose, let us 
consider the following Lagrangian of the form
\begin {equation}
L_i(x, \dot{x}) = f(x) \; \dot{x}^i - V(x) \; ,
\end {equation}
where $f(x)$ is an arbitrary function and $i$ an integer number 
different from $0$. The usual Euler-Lagrange equation leads to
\begin {equation}
(i-1)\left[ i \; f \; \dot{x}^{i-2} \ddot{x} + 
          {df \over dx}\dot{x}^{i} \right] + {dV \over dx} = 0 \; .
\end {equation}
The conjugate momentum is given by
\begin {equation}
P_i = {\partial L_i \over \partial \dot{x}} 
    = i \; f\; \dot{x}^{i-1} \; .
\end {equation}
The resulting Hamiltonian is
\begin {equation}
H_i(x, P_i) = P_i \; \dot{x} - L_i
            = {i-1 \over i}
              \left({1 \over i \; f}\right)^{1 \over i-1} 
              P_{i}^{i \over i-1}  + V(x) \; .
\end {equation}
The presence of $(i-1)$ in the denominator indicates clearly 
that the case $i=1$ requires a particular treatment. 
For $i \ne 1$, we can check that the canonical equations  
\begin {equation}
\dot{x} = { \partial H_i \over \partial P_i}
        = \left({P_i \over i \; f }\right)^{1 \over i-1}\; ,
\end {equation}
\begin {equation}
\dot{P_i} = -{ \partial H_i \over \partial x}  
          = \left({P_i \over i \; f}\right)^{i \over i-1} 
            {df \over dx} - {dV \over dx} \; ,
\end {equation}
are compatible with (73) and (74). For $i=1$, if we apply naively 
the well-known relations
\begin {equation}
P_1 = {\partial L_1 \over \partial \dot{x}}  
    =  f(x) 
\end {equation}
and
\begin {equation}
H_1(x, P_1) = P_1 \; \dot{x} - L_1
            = V(x) \; ,
\end {equation}
the canonical equations
\begin {equation}
\dot{x} = { \partial H_1 \over \partial P_1} = 0 
\end {equation}
and
\begin {equation}
\dot{P_1} = -{ \partial H_1 \over \partial x} =  - {dV \over dx} 
\end {equation}
are not compatible with (73) and (74) for $i=1$. In fact, (73) 
indicates that $dV / dx = 0$ while (78) and (81) imply that 
$dV / dx = - \dot{f}$. Furthermore, (78) indicates that $P_1$ 
and $x$ are not independent variables. The solution to this 
problem consists in adding to the Lagrangian (72) a quadratic 
term,
\begin {equation}
L_1(x, \dot{x}) = {1 \over 2} \lambda \; \dot{x}^2 
                  + f(x) \; \dot{x} - V(x) \; ,
\end {equation}
$\lambda$ being a constant, and in taking the limit 
$\lambda \to 0$ after having obtained the equation of motion. 
In the Hamiltonian formulation, it is necessary to keep 
$\lambda$ until $P_1$ and $\dot{P}_1$ are eliminated from 
the canonical equations. In this way, we can check that we 
obtain the same equation of motion with the two formulations.

\vskip\baselineskip
\noindent
{\bf Appendix II}
\vskip\baselineskip

Our goal here is to deduce the generalized canonical equations 
from the Hamiltonian (47) and to show that the Lagrangian 
formulation and the Hamiltonian one are equivalent. 

By applying relations (18)-(23), the canonical equations are 
\begin {eqnarray}
\dot{x} & \equiv & { \partial H \over \partial P} = \dot{x}\; , \\
\ddot{x} & \equiv & { \partial H \over \partial \Pi} =  \ddot{x} \; , \\
\dot{\ddot{x}} & \equiv & { \partial H \over \partial \Xi} 
                 = {1 \over  \lambda } 
     \left[
          \Xi - \sum_{n=0}^{\infty} \sum_{k=0}^{\infty} 
          { \hbar^{n} \beta_{nk} \over \mu^{n-1}} 
          {x^k \ddot{x}^{n+k-2} \over \dot{x}^{3n+2k-3}}
    \right]   \; , \\
\dot{P} & \equiv & - { \partial H \over \partial x} 
          = \sum_{n=0}^{\infty} \sum_{k=0}^{\infty} 
          { \hbar^{n} (k+1) \alpha_{n,k+1} \over \mu^{n-1}}
          {x^k \ddot{x}^{n+k+1} \over \dot{x}^{3n+2k}}
  \nonumber\\            
     &  &    \hskip10mm  + {1 \over  \lambda } 
     \left[
          \Xi - \sum_{n=0}^{\infty} \sum_{k=0}^{\infty} 
          { \hbar^{n} \beta_{nk} \over \mu^{n-1}} 
          {x^k \ddot{x}^{n+k-2} \over \dot{x}^{3n+2k-3}}
    \right] \nonumber\\            
     &  & 
    \hskip16mm \left[
          \sum_{p=0}^{\infty} \sum_{j=0}^{\infty} 
          { \hbar^{p} (j+1) \beta_{p,j+1} \over \mu^{p-1}} 
          {x^j \ddot{x}^{p+j-1} \over \dot{x}^{3p+2j-1}}
    \right] 
          \nonumber\\            
     &  &   \hskip10mm - {dV \over dx} \; , \\
\dot{\Pi} & \equiv & -{ \partial H \over \partial \dot{x}} 
          = -P 
          - \sum_{n=0}^{\infty} \sum_{k=0}^{\infty} 
          { \hbar^{n} (3n+2k-2) \alpha_{nk} \over \mu^{n-1}}
          {x^k \ddot{x}^{n+k} \over \dot{x}^{3n+2k-1}}
  \nonumber\\            
     &  &    \hskip10mm  - {1 \over  \lambda } 
     \left[
          \Xi - \sum_{n=0}^{\infty} \sum_{k=0}^{\infty} 
          { \hbar^{n} \beta_{nk} \over \mu^{n-1}} 
          {x^k \ddot{x}^{n+k-2} \over \dot{x}^{3n+2k-3}}
    \right] \nonumber\\            
     &  & 
    \hskip16mm \left[
          \sum_{p=0}^{\infty} \sum_{j=0}^{\infty} 
          { \hbar^{p} (3p+2j-3) \beta_{pj} \over \mu^{p-1}} 
          {x^j \ddot{x}^{p+j-2} \over \dot{x}^{3p+2j-2}}
    \right] 
               \; , \\
\dot{\Xi} & \equiv & -{ \partial H \over \partial \ddot{x}} 
          = -\Pi + \sum_{n=0}^{\infty} \sum_{k=0}^{\infty} 
          { \hbar^{n} (n+k) \alpha_{nk} \over \mu^{n-1}}
          {x^k \ddot{x}^{n+k-1} \over \dot{x}^{3n+2k-2}}
  \nonumber\\            
     &  &    \hskip10mm  + {1 \over  \lambda } 
     \left[
          \Xi - \sum_{n=0}^{\infty} \sum_{k=0}^{\infty} 
          { \hbar^{n} \beta_{nk} \over \mu^{n-1}} 
          {x^k \ddot{x}^{n+k-2} \over \dot{x}^{3n+2k-3}}
    \right] \nonumber\\            
     &  & 
    \hskip16mm \left[
          \sum_{p=0}^{\infty} \sum_{j=0}^{\infty} 
          { \hbar^{p} (p+j-2) \beta_{pj} \over \mu^{p-1}} 
          {x^j \ddot{x}^{p+j-3} \over \dot{x}^{3p+2j-3}}
    \right] 
               \; . 
\end {eqnarray}

Firstly, as we have indicated at the end of Section 2, Eqs. (18) 
and (20), now become (83) and (84), represent identities and do 
not give any information about the motion. 

Secondly, remark that (85) is equivalent to (46). 

Thirdly, if we calculate the temporal derivative of the two 
members of (85), we can deduce that 
\begin {eqnarray}
\dot{\Xi} = \sum_{n=0}^{\infty} \sum_{k=0}^{\infty} 
            { \hbar^{n} \over \mu^{n-1}} 
\left\lbrace 
  \left[ 
           -(3n+2k-3) \beta_{nk} + (k+1) \beta_{n,k+1} 
  \right]
           {x^k \ddot{x}^{n+k-1} \over \dot{x}^{3n+2k-2}}
\right.
      \hskip10mm && \nonumber\\  
\left. 
       +(n+k-2) \beta_{nk} {x^k \ddot{x}^{n+k-3} \dot{\ddot{x}} 
       \over \dot{x}^{3n+2k-3}}
\right\rbrace
      +\lambda \ddot{\ddot{x}} \; . \nonumber
\end {eqnarray}
Then, by substituting in this last relation $\dot{\Xi}$ by 
its expression (88) and taking into account relation (85), 
we reproduce for $\Pi$ the same expression as the one given 
by (45). 

Fourthly, if we calculate the temporal derivative of the obtained 
result for $\Pi$ (or of (45)), we get
\begin {eqnarray}
\dot{\Pi}  & = &  \sum_{n=0}^{\infty} \sum_{k=0}^{\infty} 
            { \hbar^{n} \over \mu^{n-1}} 
\left\lbrace \hskip-4.5mm \phantom{Q \over M}  \right. \nonumber \\ 
           &   &   
   \left[   \hskip-5mm \phantom{Q \over M}    
       -(3n+2k-2)[(n+k) \alpha_{nk} + (3n+2k-3) \beta_{nk}] + (k+1) [ (n 
   \right. 
       \nonumber \\ 
           &   &   
       \hskip0mm  
       +k+1) \alpha_{n,k+1} 
   \left.  
       \hskip0mm + (6n+4k-3) \beta_{n,k+1} - (k+2) \beta_{n,k+2} ]
       \hskip-4.5mm \phantom{Q \over M}
   \right]
       {x^k \ddot{x}^{n+k} \over \dot{x}^{3n+2k-1}}  
       \nonumber \\ 
           &   &  
       + (n+k-1)
   \left[ \hskip-4.5mm \phantom{Q \over M}      
       (n+k) \alpha_{nk} + (3n+2k-3) \beta_{nk} 
   \right.
       \nonumber \\ 
           &   &
   \left.    
      \hskip48mm - (k+1) \beta_{n,k+1} 
       \hskip-4.5mm \phantom{Q \over M}
   \right]
       {x^k \ddot{x}^{n+k-2} \dot{\ddot{x}} \over \dot{x}^{3n+2k-2}}  
\left. \hskip-4.5mm \phantom{Q \over M} \right\rbrace  \nonumber \\ 
    &   &
     - \lambda \;\dotFive{x} \; . \nonumber
\end {eqnarray}
Then, by substituting this expression in (87) and taking into 
account relation (85), we reproduce for $P$ the same expression 
as the one given by (44). 

It follows that the canonical equations (85), (87) and (88) 
lead to the same expressions (44), (45) and (46) for the 
conjugate momentums $P$, $\Pi$ and $\Xi$ obtained from the 
Lagrangian (43). 

Fifthly, from (43) and (47), we can easily check that 
\begin {equation}
{\partial L \over \partial x} = - {\partial H \over \partial x} \; ,
\end {equation}
where we have used (85). Thus, since we have obtained the same 
expression for $P$ in the two approaches, by comparing (11) 
and (19), we deduce that the Hamiltonian formulation is 
equivalent to the Lagrangian one and both of them lead to 
the same law of motion for any $\lambda$. 

\vskip\baselineskip
\noindent
{\bf Appendix III}
\vskip\baselineskip

In this Appendix, we will search for the numerical values of 
$\alpha_{nk}$ and $\beta_{nk}$ with which relation (57) is valid.
For this purpose, let us calculate the second and the third 
derivatives of $S_0$ from (52). We obtain
\begin {eqnarray}
{d^{2}S_0 \over dx^{2}} & = &   
                 \sum_{n=0}^{\infty} \sum_{k=0}^{\infty} 
                 { \hbar^{n} \over \mu^{n-1}} 
                 \left\lbrace 
                 \phantom{ Q \over M   } 
                 \right.  \nonumber\\  
  &  & [(k+1) A_{n,k+1} - (3n+2k-1) A_{nk}] 
        {x^k \ddot{x}^{n+k+1} \over \dot{x}^{3n+2k+1}}
       \nonumber\\ 
  &  & + [(n+k) A_{nk} - (3n+2k-2) B_{nk} + (k+1)B_{n,k+1}] 
        {x^k \ddot{x}^{n+k-1} \dot{\ddot{x}} \over \dot{x}^{3n+2k}}  
       \nonumber\\
  &  & + (n+k-2) B_{nk}
       {x^k \ddot{x}^{n+k-3} \dot{\ddot{x}}^2 \over \dot{x}^{3n+2k-1}}
        \left. 
            + B_{nk}
            {x^k\ddot{x}^{n+k-2} \ddot{\ddot{x}} \over \dot{x}^{3n+2k-1}}
       \right\rbrace 
    \; ,
\end {eqnarray}
and
\begin {eqnarray}
{d^{3}S_0 \over dx^{3}} & = & 
      \sum_{n=0}^{\infty} \sum_{k=0}^{\infty} 
       { \hbar^{n} \over \mu^{n-1}} 
       \left\lbrace 
       \phantom{  Q \over M } 
       \right. \nonumber\\
  &  & [(3n+2k-1)(3n+2k+1)A_{nk}-2(k+1)(3n+2k+1)A_{n,k+1} \nonumber\\ 
  &  &
      \hskip45mm + (k+1)(k+2)A_{n,k+2} ]
       {x^k \ddot{x}^{n+k+2} \over \dot{x}^{3n+2k+3}} \nonumber\\ 
  &  & +[- (6n^2+4k^2+10nk+2n+k-1)A_{nk} \nonumber\\ 
  &  & + (3n+2k)(3n+2k-2)B_{nk} + 2(k+1)(n+k+1)A_{n,k+1} \nonumber\\ 
  &  & \hskip0mm  -2(k+1)(3n+2k)B_{n,k+1} + (k+1)(k+2)B_{n,k+2}]
       {x^k \ddot{x}^{n+k} \dot{\ddot{x}} \over \dot{x}^{3n+2k+2}} 
       \nonumber\\ 
  &  & +[ (n+k)(n+k-1)A_{nk} \nonumber\\ 
  &  & \hskip20mm -(6n^2+4k^2+10nk-12n-9k+4)B_{nk} \nonumber\\ 
  &  & \hskip34mm + 2(k+1)(n+k-1)B_{n,k+1} ]
       {x^k \ddot{x}^{n+k-2} \dot{\ddot{x}}^2 \over \dot{x}^{3n+2k+1}} 
       \nonumber\\ 
  &  & +(n+k-2)(n+k-3) B_{nk}  
       {x^k \ddot{x}^{n+k-4} \dot{\ddot{x}}^3 \over \dot{x}^{3n+2k}} 
       \nonumber\\
  &  & +[(n+k) A_{nk}-(6n+4k-3) B_{nk} + 2(k+1) B_{n,k+1}]   
       {x^k \ddot{x}^{n+k-1} \ddot{\ddot{x}} \over \dot{x}^{3n+2k+1}} 
       \nonumber\\ 
  &  & \left. 
       +3(n+k-2) B_{nk}
       {x^k \ddot{x}^{n+k-3} \dot{\ddot{x}}\ddot{\ddot{x}} 
       \over \dot{x}^{3n+2k}}
       + B_{nk}
       {x^k \ddot{x}^{n+k-2} \dotFive{x} \over \dot{x}^{3n+2k}}
   \right\rbrace  \; .
\end {eqnarray}
The values of $\alpha_{nk}$ and $\beta_{nk}$ can be determined 
by writing the two members of (57) as a power series with 
respect to $\hbar$ by using (52), (90) and (91). At the classical 
level $(\hbar=0)$, Eqs. (52) and (57)  can be written as   
\begin {equation}
\left({dS_0 \over dx}\right)^{(0)} =    
       \sum_{k=0}^{\infty}   \mu      
    \left[
         A_{0k} {x^k \ddot{x}^{k} \over \dot{x}^{2k-1}} 
         +  B_{0k} {x^k \ddot{x}^{k-2} \dot{\ddot{x}} 
           \over \dot{x}^{2k-2}}  
    \right] 
     \; ,
\end {equation}
and
\begin {equation}
      \dot{x} \left({dS_0 \over dx}\right)^{(0)} - 
      { 1 \over 2\mu} \left[\left({dS_0 \over dx}\right)^{(0)}\right]^2 
     - \sum_{k=0}^{\infty} 
            \mu 
    \left[
         \alpha_{0k} {x^k \ddot{x}^{k} \over \dot{x}^{2k-2}} 
         + \beta_{0k} {x^k \ddot{x}^{k-2} \dot{\ddot{x}} 
           \over \dot{x}^{2k-3}}  
    \right] 
    = 0  
  \; .
\end {equation}
The terms proportional to $\dot{\ddot{x}}^2$ that we obtain 
by substituting  (92) in (93) are linearly independent. Thus, 
Eq. (93) can not be satisfied without imposing any condition on 
the particle motion unless one has  
\begin {equation} 
B_{0k} = 0, \ \ \ \ \forall \ \ k \geq 0  \; .
\end {equation}
By using this result in (92), relation (93) turns out to be
\begin {equation}
     \sum_{k=0}^{\infty} A_{0k} {x^k \ddot{x}^{k} \over \dot{x}^{2k-2}}
     -  { 1 \over 2} 
     \left[  
          \sum_{k=0}^{\infty} A_{0k} {x^k \ddot{x}^{k} 
          \over \dot{x}^{2k-1}} 
     \right]^2 
         - \sum_{k=0}^{\infty} 
     \left[
         \alpha_{0k} {x^k \ddot{x}^{k} \over \dot{x}^{2k-2}} 
         + \beta_{0k} {x^k \ddot{x}^{k-2} \dot{\ddot{x}} 
           \over \dot{x}^{2k-3}}  
     \right] 
     = 0  
  \; .
\end {equation}
The terms proportional to $\dot{\ddot{x}}$ are linearly independent. 
Thus, as above, (95) can not be satisfied unless one has
\begin {equation} 
\beta_{0k} = 0, \ \ \ \ \forall \ \ k \geq 0  \; .
\end {equation}
Using expression (54) for $n=0$ and taking into account (94) 
and (96), we deduce that 
$k(k-1) \alpha_{0k} = 0 \ \; \forall \ k \geq 0$. 
It follows that $\alpha_{0k} = 0 \ \; \forall \ k \geq 2$. 
Therefore, with the use of (53) for $n=0$ and (96), Eq. (95) 
becomes
\begin {eqnarray}
      \left(-2 \alpha_{00}^2 -{1 \over 2} \alpha_{01}^2 + 
      2 \alpha_{00} \alpha_{01} + \alpha_{00} 
      -\alpha_{01}\right) \dot{x}^2         \hskip20mm&& \nonumber\\    
      -{1 \over 2} \alpha_{01}^2 {x^2 \ddot{x}^2 \over \dot{x}^2  } 
     - \alpha_{01} (2\alpha_{00} - \alpha_{01})  x \ddot{x} = 0 \; .
\end {eqnarray}
In this relation, we have three terms linearly independent. It 
follows that $\alpha_{01}=0$ and then 
$\alpha_{00}(1 - 2 \alpha_{00})=0$. The solution $\alpha_{00}=0$ 
corresponds to a trivial solution and does not allow to obtain 
the classical limit. We keep the other solution and we write
\begin {equation} 
    \alpha_{00}= {1 \over 2}, \ \ \ \ \ \ \alpha_{0k} = 0, 
     \ \ \ \ \forall \ k \geq 1  \; .
\end {equation}

Note that with the above results, we have $A_{00}=1$ and 
$A_{0k}=0 \ \forall \ k \geq 1$. Then, by taking into account 
relation (94), expression (92) becomes 
\begin {equation}
\left({dS_0 \over dx}\right)^{(0)} = \mu  \dot{x}         \; .
\end {equation}

At the first level, taking into account relations (96) and 
(98), Eq. (57) yields
\[
 \hbar \mu^{2} \sum_{k=0}^{\infty}
    \left[
          \alpha_{1k}  {x^k \ddot{x}^{k+1} \over \dot{x}^{2k-1}} +
          \beta_{1k}  {x^k  \ddot{x}^{k-1} \dot{\ddot{x}} 
           \over {\dot{x}^{2k-2} }}
    \right] 
    = 0 \; . 
\]
In this equality, all the terms are linearly independent. If
we do not want to impose any condition on the particle motion, 
it is necessary that
\begin {equation}
    \alpha_{1k} =  0, \ \ \ \ \ \ \ \ \ \beta_{1k} = 0, 
    \ \ \ \forall \ k \geq 0 \; .
\end {equation}
These relations mean that $A_{1k}=0$ and 
$B_{1k}=0 \ \forall \ k \geq 0$ and that the derivatives 
(52), (90) and (91) of $S_0$ do not contain terms 
proportional to $\hbar$. Then, at the second level, from 
(57) we have
\[
      \hbar^{2} \mu
\left[
     \left(\alpha_{20} -{5 \over 8}\right) {\ddot{x}^2 \over \dot{x}^{2}}
     + \left(\beta_{20} +{1 \over 4}\right) {\dot{\ddot{x}} \over \dot{x}}
     + \sum_{k=1}^{\infty}
     \left(
           \alpha_{2k} {x^k \ddot{x}^{k+2} \over \dot{x}^{2k+2}} 
           + \beta_{2k} {x^k \ddot{x}^k \dot{\ddot{x}} 
             \over \dot{x}^{2k+1}}
     \right) 
\right] 
    = 0 \; .
\]
For the same reasons as above, we deduce that
\begin {equation}
    \alpha_{20} =  {5 \over 8}, \ \ \ \ \ \ \beta_{20} = -{1 \over 4}, 
    \ \ \ \ \ \ \alpha_{2k} = 0 , \ \ \ \ \ \ \beta_{2k} =0, 
    \ \ \ \ \ \  \forall \ k \geq 1  \; .
\end {equation}
In order to determine the other parameters $\alpha_{nk}$ and  
$\beta_{nk}$ for $n>2$, it is essential to remark that with 
the values given in (101), we can see from (53) and (54) that 
$A_{2k}=0$ and  $B_{2k}=0$ for every $k$ even for $k=0$.  
Thus, as it is the case for the terms proportional to $\hbar$, 
the derivatives (52), (90) and (91) of 
$S_0$ do not contain terms proportional to $\hbar^2$.  
For the upper levels, we will use a reasoning by recursion. 
Let us begin by the third level. Taking into account relations 
(96), (98), (100) and (101), it is easy to check that the 
validity of (57) requires that 
\[
      \hbar^{3} \; \sum_{k=0}^{\infty}
\left[
           \alpha_{3k} {x^k \ddot{x}^{k+3} \over \dot{x}^{2k+5}} 
           + \beta_{3k} {x^k \ddot{x}^{k+1} \dot{\ddot{x}} 
             \over \dot{x}^{2k+4}}
\right] 
    = 0 \; ,
\]
which leads to
\begin {equation}
    \alpha_{3k} =  0, \ \ \ \ \ \ \ \beta_{3k} = 0, 
     \ \ \ \ \ \ \ \forall \ k \geq 0 \; .
\end {equation}
Now, let us suppose that for $n>3$, the validity of (57) 
at the $(n-1)^{th}$ requires that $\alpha_{3k}=0$, ..., 
$\alpha_{n-1,k}=0$ and $\beta_{3k}=0$, ..., $\beta_{n-1,k}=0$ 
for every $k \geq 0$. We have then $A_{3k}=0$, ..., 
$A_{n-1,k}=0$ and  $B_{3k}=0$, ..., $B_{n-1,k}=0$ for every 
$k \geq 0$. By taking into account the fact that $A_{00}=1$, 
$A_{0k}=0$ for every $k \geq 1$ and 
$B_{0k}=A_{1k}=B_{1k}=A_{2k}=B_{2k}=0$ for every $k \geq 0$, at 
the $n^{th}$ level, relation (57) yields 
\[  
   { \hbar^{n} \over \mu^{n-3}}
    \sum_{k=0}^{\infty}
     \left[
           \alpha_{nk} {x^k \ddot{x}^{n+k} \over \dot{x}^{3n+2k-4}} 
           + \beta_{nk} {x^k \ddot{x}^{n+k-2} \dot{\ddot{x}} 
             \over \dot{x}^{3n+2k-5}}
     \right]  
    = 0 \; .   
\]
As above, this relation implies that $\alpha_{nk} =0$ and 
$\beta_{nk}=0$ $\forall \ k \geq 0$. Now we can assert that, 
in addition to the conditions (96), (98), (100) and (101), the 
validity of (57) requires that 
\begin {equation}
      \alpha_{nk} = 0 , \ \ \ \ \beta_{nk} = 0 , 
      \ \ \ \ \forall \ \ \ n \geq 3 , \ \ \ \forall \ \ \ k \geq 0 \;. 
\end {equation}
%
%

\vskip\baselineskip
\noindent
{\bf References }

\end {document}